# DEVELOPMENT OF A COMPACT PHOTO-INJECTOR WITH RF-FOCUSING LENS FOR SHORT PULSE ELECTRON SOURCE APPLICATION

[1]Alexander Grabenhofer, [2]Douglas W. Eaton, and [1,3]Young-Min Shin
[1]Department of Physics, Northern Illinois University, Dekalb, IL, 60115, USA
[2]ScandiNova Systems AB, Uppsala, Sweden
[3]Fermi National Accelerator Laboratory (FNAL), Batavia, IL 60510, USA

*Abstract*

For development of compact ultrafast electron source system, we are currently designing a short-pulse RF-gun with RF focusing structure by means of a series of comprehensive modeling analysis processes. EM design of a 2.5 cell resonant cavity with input coupler, acceleration dynamics of photo-emitted electron bunch, EM design of RF-lens with input coupler, and phase-space analysis of focused electron bunch are systematically examined with multi-physics simulators. All the features of the 2.856 GHz cavity geometry were precisely engineered for acceleration energies ranging from 100 keV to 500 keV (safety limited) to be powered by our 5 MW S-band klystron. The klystron (Thales TH2163) and modulator system (ScandiNova K1 turnkey system) were successfully installed and tested. Performance tests of the klystron system show peak output power > 5 MW, as per operation specifications. At the quasi-relativistic energies, the electron source is capable of generating 100fC – 1 pC electron bunch with pulse duration close to 30 fs – 1 ps and transverse size of a few hundred microns. PIC simulations have shown that the electron bunch undergoes fast RF acceleration, rapidly reaching the desired energies, which can be controlled by tuning RF injection phase and input driving power. It has been shown that it is possible to also focus/compress the bunch longitudinally using a RF-lens, which would allow us to control the temporal resolution of the system as well. While our primary analysis has been performed on a 2.5 cell design, we are also looking into half-cell (single cavity) design that is expected to provide the same range of beam energy with a simple configuration.

## INTRODUCTION

The ultrafast electron diffraction (UED) source is one of the most powerful time-resolved spectroscopic/ microscopic tools capable of visualizing the sub-atomic world in a femto-second time frame. A successfully developed UED system will enable us to observe various ultrafast phenomena in chemical and biological substances of solid, gas and liquid phases. The instrument will be a useful tool for a wide range of scientific experiments in chemistry, biology, physics, radiolysis, and engineering on which accurate spatiotemporal analysis methodology is highly demanded. Ultrafast technology with sub-atomic resolution within the femto-second range, while still immature, far exceeds the current technical capacity of commercially available electron microscopes. This technology is not achievable from any single unit or commercial product, as integral parts and their systematic integration are still under development. For spectroscopic application, the main advantage of electron diffraction is that the electrons scatter off all atoms and atom-atom pairs in the molecular sample. Thus, unlike spectroscopy wherein the probe is turned to specific transitions, the electron probe is sensitive to all species in its path and can hence uncover structures that spectroscopy may be blind to. This type of spectroscopic imaging technique can offer spatiotemporal information of transitional dynamics and chemical reaction in liquid and solid states that are of main interest in NIU, as described in the next section, and the proposed UED system is expected to secure sufficient spatial and temporal resolution with good spatial coherence for those kinds of studies. The UED technique employs time-delayed ultrafast pulses – a laser pulse to initiate the reaction and an electron pulse to probe the ensuing structural change in the molecular sample. The resulting electron diffraction patterns are then recorded on a CCD camera. These processes can only produce femto-second electron diffraction when all the elements are properly orchestrated in harmony, which requires a sophisticated design process. In particular, as a key element of a UED system, a short pulse electron source must be capable of producing highly mono-energetic electron bunches of a few hundred keV to a few MeV to create dynamic scattering images with femto-second/sub-angstrom spatiotemporal resolution. Currently, we have a 5.5 MW S-band klystron driver for a photo-injector that has been designed with a 2.5 cell RF-gun to produce 100 ~ 500 keV electron bunches.

## BACKGROUND AND OBJECTIVES

Most of UED systems have employed the photocathode-based DC gun with an energy in the range of 30 ~ 100keV (Max. DC field = 10 ~ 12 MV/m) and 400 ~ 600 fs at $10^3$~$10^4$ e-/pulse. However, with the DC guns, bunch length and energy spread are increased during beam transport. The bunch length of a 30 keV e-beam is, for instance, increased from fs to a few ps and the energy spread ($\Delta E/E$) becomes ~ $3 \times 10^{-3}$ to a distance of 40 cm. It is thus difficult to obtain a < 100 fs electron bunch with $\Delta E/E < 10^{-3}$ using DC guns. Also, to reduce the space charge effect in low energy DC gun system, the distance between sample and cathode should be minimized (4 ~ 5 cm) and/or the number of electrons in bunch should be decreased to $10^3$ e-/bunch with 400 fs. It is difficult to observe the ultrafast dynamics with single-shot measurement and the studies of low-energy UED are limited to the reversible processes. Recently, photocathode RF-guns with higher accelerating gradients and electron density are therefore widely investigated for femto-second UED system application, which is also considered in our approach. We plan to pursue mechanical design and manufacturing of a RF-gun and high vacuum chambers, including a cooling system. The machined parts will be inspected by the Engineering Department and our HP8510C VNA will be available for spectrum analysis of the manufactured RF cavity.

## SYSTEM CONFIGURATION

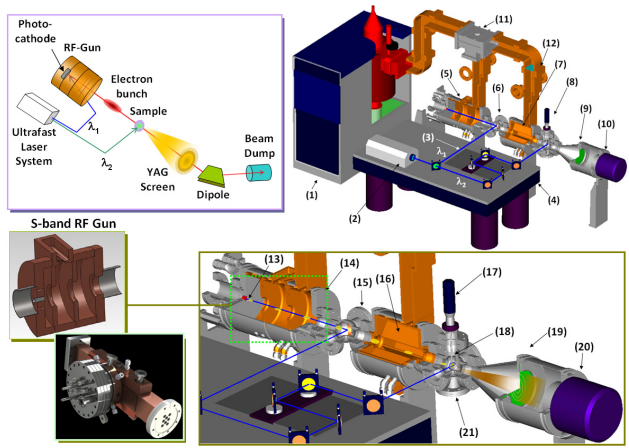

Figure 1: (a) Conceptual sketch of ultrafast electron diffraction system (b) engineering drawing of the designed NIU UED system. The RF power source (S-band klystron system) is already under installation with several other RF components (insets: designed 2.5 cell cavity and S-band photoinjector, courtesy of Radiabeam). See text for details.

Figure 1 is an engineering sketch of the proposed instrument that mainly consists of ultrafast laser system, a RF-gun, beam focusing devices, a 6-way cross sample chamber with a micrometer aligner, Yttrium aluminium garnet (YAG) screen with femto-second optoelectronic CCD camera, and electron beam dump with a bending section, if necessary. The electron bunch is photo-emitted by illuminating a photocathode with an ultra-short laser (that could eventually be spatiotemporally shaped). This pump-probe configuration, where the pump is a μJ laser pulse and the probe is the electron beam, could have enormous advantages in studying the dynamics of system in transient states with sub 100-fs (and eventually attosecond) time resolution. The short pulse electron source in the proposed UED unit mainly includes a femto-second laser system, a RF-gun, and a buncher for pulsed operation with femto-second electron packets. The system will employ a regeneratively mode-locked laser that produces ultra-short pulses of up to 10 ~ 100 nJ at 800 nm with variable pulse width (100 fs to 10 ps) and repetition rate (~ 80 MHz). The output pulses pass through two successive nonlinear crystals to be frequency tripled (267 nm). The harmonic is separated from the residual infrared radiation (IR) beam by dichroic mirrors, and the frequency-tripled pulses are introduced to the photocathode of the RF-gun for generating the electron pulse. The residual IR fundamental beam remains available to heat or excite samples and clock the time through a computer-controlled optical delay line for time-resolved applications. In order to pursue electron diffraction spectroscopy with sufficiently high signal-to-noise ratio, electron bunches need to contain $> 10^6$ electrons per a pulse. A photocathode with a quantum efficiency of $10^{-6}$ will require $10^{12}$ UV photons per a pulse to provide the $10^6$ electrons/pulse. 10% efficiency in the two harmonic generation steps from the near IR to the UV implies $10^{14}$ IR photons/pulse. For the ~ 1eV IR photon energy, one then arrives at an IR laser pulse energy of ~ 10 μJ. A 10 μJ laser pulse energy is difficult to obtain from a fs laser oscillator directly as 0.1 – 1 μJ is about the limit with long cavity and/or cavity dumping techniques. A high QE ($> ~ 10^{-4}$) photocathode may thus be needed in order to create $> 10^6$ electrons/pulse with the laser energy. The first approach under our consideration is to use a Magnesium (Mg) photocathode (work function: 3.66 eV) at 267nm that has demonstrated relative high QE ($10^{-3}$) under modest vacuum condition [1].

## SIMULATION MODELING ANALYSIS

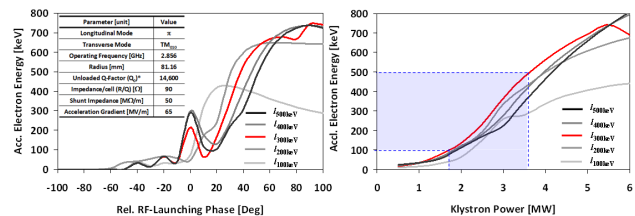

Figure 2: Optimally designed RF-gun parameters (table) and RF-launching phase and klystron power versus electron energy graphs obtained from tracking/PIC simulations with various cell-to-cell distances for beam-wave synchronization (1 pC bunch charge and 100 fs pulse).

As shown in Fig. 2, the acceleration performance of the design RF gun was examined with CST PIC solver. CST PIC code is the 3D particle-in-cell simulator based on finite-difference-time-domain (FDTD) algorithm that includes space charge effects of an electron bunch, represented by a collection of macro-particles, using real-time field profiles and ambient energy losses. The macro-particles are deposited on a grid and Poisson's equation is solved in the rest frame of the electron bunch and the electrostatic fields are Lorentz-transformed to compute the total force experienced by each macro-particle in laboratory frame. For the particle simulation, the electric and magnetic fields of 2.856 GHz, π, $TM_{010}$-mode, provided by CST Eigenmode solver, are imported into the cavity in particle-tracking and PIC simulation models. The initial laser distribution was taken to be an ultra-short laser pulse and a perfect three dimensional ellipsoidal distribution. At quasi-relativistic energies the electron source is capable of generating 100fC ~ 1 pC electron bunch with pulse duration close to 30 fs ~ 1 ps and transverse size of few hundred microns. At lower energies similar performances can be achieved with lower (100 fC) bunch charge. The space charge effect appears more dominant at the lower energies in that transverse and longitudinal emittances are diluted in the phase-space. A preliminary design is examined with the RF-gun by particle-tracking and PIC simulations. The electron bunch emitted from the cathode is slightly blown up for Columbic space charge force and then rapidly decreases toward the gun exit, but once entering in the focusing field region, it is rapidly squeezed in the axial directions. The simulations show that the bunch has improved momentum linearization in phase-space with spatial compression.

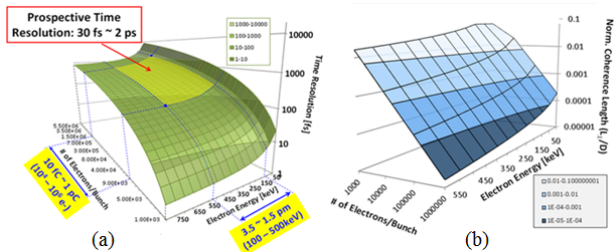

Figure 3: (a) spatiotemporal resolution and (b) coherence length versus # of electrons and electron energy.

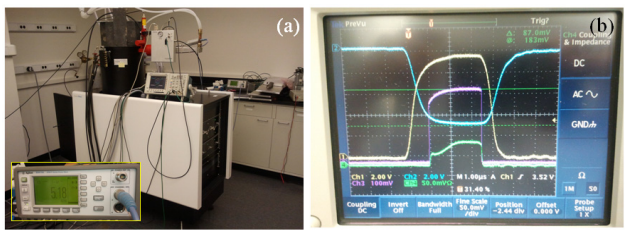

Figure 4: (a) Assembled S-band klystron (Thales TH2163) and modulator system (inset: output power measured by a power-meter) (b) waveforms of pulsed power measured by an oscilloscope.

A crucial component in the case of pump-probe experiment will be the precise measurement of the delay time between the laser pump and electron probe beams. Time stamping with 10 fs resolution will be achieved using an electro-optical modulator [2,3]. If the device does not offer fast enough response to distinct temporal evolution of diffracting patterns, it may be combined with streak camera to improve time resolution, which may also require a deflecting device. The time-resolution of the best optoelectronic streak cameras is around 100 femto-seconds. Measurement of pulses shorter than this duration requires other techniques such as optical autocorrelation and frequency-resolved optical gating (FROG) [4,5]. Figure 3 shows prospective specification of our designed UED system that is assessed by analytic calculations [6].

One can see in Fig. 3 that the 100 ~ 500 keV, corresponding to 1.5 ~ 3.5 pm spatial resolution, leads to 30 fs ~ 2 ps time resolution with ~ $10^4$ ~ $10^6$ electrons (~ 10 fC ~ 1 pC).

## RF-TEST OF KLYSTRON SYSTEM

As a part of plan of a short pulse electron beam source construction, we installed and commissioned the S-band klystron (Thales TH2163) and modulator system (ScandiNova K1 turnkey system), as shown in Fig. 4. The high power RF-test ended up successfully demonstrating a ~ 5.2 MW peak power with 4 μs pulse width. The optical table and some waveguide components were added to the system, which will be used as a high power driver for the RF-gun once an accelerating mode cavity is added.

## CONCLUSION

We have been developing a short pulse electron source for ultra-fast electron diffraction (UED) application at Northern Illinois University (NIU). The construction of the compact ultrafast electron source and associated detection equipment will be a great opportunity for the unique resource and technology in beam and laser physics, electrical engineering and other associated accelerator technologies, and support users from biology, chemistry, condensed matter physics, and medicine, among other fields.